# The surface-forming energy release rate based fracture criterion for elastic-plastic crack propagation

S. Xiao, H.L. Wang[*], B. Liu[*]

*AML, CNMM, Department of Engineering Mechanics, Tsinghua University, Beijing 100084, China*

[*]Corresponding authors. Tel.: 86-10-62786194; fax: 86-10-62781824

E-mail address: liubin@tsinghua.edu.cn (B. Liu);
wang-hl@mails.tsinghua.edu.cn(H.-L. Wang)

**Abstract**

J integral based criterion is widely used in elastic-plastic fracture mechanics. However, it is not rigorously applicable when plastic unloading appears during crack propagation. One difficulty is that the energy density with plastic unloading in J integral cannot be defined unambiguously. In this paper, we alternatively start from the analysis on the power balance, and propose a surface-forming energy release rate (ERR), which represents the energy directly dissipated on the surface-forming during the crack propagation and excludes the loading-mode-dependent plastic dissipation. Therefore the surface-forming ERR based fracture criterion has wider applicability, including elastic-plastic crack propagation problems. Several formulae have been derived for calculating the surface-forming ERR. From the most concise formula, it is interesting to note that the surface-forming ERR can be computed only by the stress and deformation of the current moment, and the definition of the energy density or work density is avoided. When an infinitesimal contour is chosen, the expression can be further simplified. For any fracture behaviors, the surface-forming ERR is proven to be path-independent, and the path-independence of its constituent term, so-called $J_s$ integral, is also investigated. The physical meanings and applicability of the proposed surface-forming ERR, traditional ERR, $J_s$ integral and $J$ integral are compared and discussed.

## 1. Introduction

Fracture is a very important problem in engineering applications. Currently, fracture mechanics has been developed relatively successfully for cracks in brittle materials (Griffith 1921, 1924; Irwin 1948; Orowan 1949; Irwin 1958) or ductile materials with only small scale plastic yielding region near a crack tip (Irwin 1958).

When the yielding region is not small, the path-independent J integral (Rice et al. 1967) can be used as a character parameter in a fracture criterion. But it is only valid when a crack does not propagate or propagates just a small amount such that no material point experiences unloading and the material can be treated as hyperelastic materials. As the crack propagates further, plastic unloading region appearing at the crack wake (as shown in Fig. 1) should not be ignored. Therefore, the J integral based criterion is no longer applicable to elastic-plastic crack propagation. Some scholars have measured a series of metals' fracture resistance curves $J_R$ at various situations. Cotterell and Atkins (1996) found that the $J_R$ curves differ from each other and are affected by constraint boundaries, because the plastic zone related to the energy dissipation is influenced by constraint boundaries. Hutchinson (1974) suggested to change the strain energy density defined in J integral to the internal energy density, and assumed that the behavior of the material is time-independent and not limited to a deformation theory of plasticity. The interpretation of J integral based on energy balance concept in the presence of irreversible plastic deformation was appraised by Sumpter and Turner (1976). Brust and Atluri (1986) adopted an incremental theory of plasticity and introduced the total accumulated increments of stress working density, then defined a path-independent integral, i.e. $T^*$ integral. They investigated the difference between J and $T^*$ resistance curves during crack initiation and after some amount of growth. When the crack finally propagates in steady-state way, $T^*$ levels out to a constant value and J continues to rise. In addition, Simba et al. (2008) discussed the J-integral and crack driving force in elastic-plastic materials, with particular emphasis on incremental plasticity. These revisions on J integral included more influence factors, but have not been widely used to predict the elastic-plastic crack propagation.

Broberg (1968) proposed another type of fracture criterion for elastic-plastic crack propagation, i.e. the essential work of fracture (EWF). According to this theory, only the work performed at the fracture process zone is a material constant, and is called the specific essential work of fracture ($w_e$) (also refer to Cotterell and Reddelt, 1976; Wnuk and Read, 1986). In addition, the specific work of fracture represents the work done in local straining, necking, and material separation near the plane of the propagating crack, and is considered to be consistent with the critical J-integral for initiation of tearing (Wnuk and Read, 1986). However, the EWF approach is not applicable to some experiments, in which unreasonable negative values of $w_e$ are obtained (Vu-Khanh, 1994).

Therefore, up to now, a widely accepted fracture criterion for elastic-plastic crack propagation still lacks. The motivation of this study is to rigorously derive a criterion which is in integral form as J integral, but only extracts the surface-forming energy release rate (not a total energy release rate) similar to the EWF criterion. The paper is structured as follows. In Section 2, starting from the balance of the power, we derive several different formulae for computing the surface-forming energy release rate under different conditions. The path-independence of the surface-forming energy release rate and its constituent terms is investigated in Section 3. In Section 4, we present the physical interpretation and advantages of the proposed fracture criterion and discuss other related issues. Main conclusions are summarized in Section 5.

## 2. The surface-forming energy release rate

Considering that the plastic unloading appears during the elastic-plastic crack propagation, the strain energy density used in traditional J integral is not clearly defined any more. Since the power always has the unambiguous meaning for any constitutive behaviors, we start our derivation from the balance of the power, which differs from previous studies.

**2.1 The power balance for crack propagation**

We study the power balance of an area (denoted by $A$) within a curve surrounding the crack tip (denoted by $\Gamma$) as shown in Fig. 2. The power used to separate the crack surface $\dot{w}_s$ can be expressed as

$$\dot{w}_s = \int_\Gamma n_j \sigma_{ij} \dot{u}_i d\Gamma - \int_A \sigma_{ij} \dot{\varepsilon}_{ij} dA \tag{1}$$

where the first part of the right-hand-side, $\int_\Gamma n_j \sigma_{ij} \dot{u}_i d\Gamma$, represents the power of the external force, and the second part, $\int_A \sigma_{ij} \dot{\varepsilon}_{ij} dA$, is the power of the internal force. $n_j$ represents the unit external normal vector of $\Gamma$, $\sigma_{ij}$, $\varepsilon_{ij}$, and $u_i$ are stress, strain and displacement components, respectively. $\dot{(\ )}$ represents the temporal derivative. We define the surface-forming energy release rate (ERR) $G_s$ by

$$\dot{w}_s = G_s \dot{a} \tag{2}$$

where $a$ is the crack length. $G_s$ represents the energy directly dissipated on the surface-forming during the crack propagation and the energy dissipation due to the plastic deformation away the surfaces is excluded. We then obtain a $G_s$ based fracture criterion

$$\begin{cases} G_s > G_{sc}, & \text{crack grows} \\ G_s < G_{sc}, & \text{no crack growth} \end{cases} \tag{3}$$

where $G_{sc}$ represents the corresponding resistance and sometimes can be regarded as a material constant, such as twice the surface energy $2\gamma$. Obviously, this criterion is suitable for elastic-plastic crack propagation and other situations.

In order to compute the surface-forming ERR $G_s$ easily, an integral form similar to J integral is needed. We introduce an accumulated work density as

$$\hat{w}(x_1, x_2, t) = \int_{t_{ref}}^{t} \sigma_{ij} \dot{\varepsilon}_{ij} d\tau + \hat{w}_{ref}(x_1, x_2) \tag{4}$$

where $t$ represents the time moment and $t_{ref}(x_1, x_2)$ is the reference moment for the point with the coordinates $(x_1, x_2)$, $\hat{w}_{ref}(x_1, x_2)$ is the corresponding accumulated work density at $t_{ref}(x_1, x_2)$. Obviously, the temporal derivative of $\hat{w}$ is

$$\frac{d\hat{w}}{dt} = \sigma_{ij} \dot{\varepsilon}_{ij} \tag{5}$$

It should be noted that Eq. (5) holds for any selection of $t_{ref}(x_1, x_2)$ and $\hat{w}_{ref}(x_1, x_2)$. In the later derivation, there are two special ways to select $t_{ref}(x_1, x_2)$ and $\hat{w}_{ref}(x_1, x_2)$:

(1) $t_{ref} = -\infty$ is chosen as the moment at which material points have no stress and have not experienced any plastic deformation, and $\hat{w}_{ref}(x_1, x_2) = 0$. The accumulated work density $\hat{w}$ under this condition is denoted as:

$$\hat{w}_{-\infty}(x_1, x_2, t) = \int_{-\infty}^{t} \sigma_{ij} \dot{\varepsilon}_{ij} d\tau \tag{6}$$

If the material is hyperelastic, $\hat{w}$ degenerates to the strain energy density $w$.

(2) Take $t_{ref}$ as the current moment $t_0$, and $\hat{w}_{ref}(x_1, x_2) = 0$, then $\hat{w}(x_1, x_2, t_0) = 0$.

In addition, if there is no special notification in later derivation, the symbol $\hat{w}$ means the arbitrary selection of $t_{ref}(x_1, x_2)$ and $\hat{w}_{ref}(x_1, x_2)$.

**2.2 Formulae of the surface-forming energy release rate for general situations**

To study the change of energy during crack propagation, a moving coordinate system ($x_1^{'}, x_2^{'}$) as shown in Fig. 3 is also used in our derivations, and its origin is always located on the moving crack tip. Correspondingly, there is a moving contour enclosing an area $A_{mov}$ besides a stationary contour enclosing an area $A_{sta}$. They are both traversed in the counterclockwise sense. At the current moment $t_0$, the stationary coordinate system ($x_1, x_2$) and the moving coordinate system ($x_1^{'}, x_2^{'}$) coincide, and the crack length is denoted by $a_0$. In this paper, only mode I crack is studied to ensure its propagation along a straight line.

At a later moment $t$, we have the relationship

$$x_1^{'}(t) = x_1 - [a(t) - a_0], \quad x_2^{'}(t) = x_2 \tag{7}$$

where $a(t)$ is the corresponding crack length.

For an arbitrary physical quantity $\Phi$, there exist the following relations between the two coordinate systems

$$\Phi = \Phi(x_1', x_2', t) = \Phi\left(x_1 - [a(t) - a_0], x_2, t\right) \tag{8}$$

$$\dot{\Phi} = \left.\frac{\partial \Phi}{\partial t}\right|_{x_1, x_2} = \left.\frac{\partial \Phi}{\partial t}\right|_{x_1', x_2} - \dot{a}\left.\frac{\partial \Phi}{\partial x_1'}\right|_{x_2, t} \tag{9}$$

Considering

$$\left.\frac{\partial \Phi}{\partial x_1'}\right|_{x_2, t} = \left.\frac{\partial \Phi}{\partial x_1}\right|_{x_2, t} \left.\frac{\partial x_1}{\partial x_1'}\right|_{x_2, t} = \left.\frac{\partial \Phi}{\partial x_1}\right|_{x_2, t} \tag{10}$$

we obtain

$$\dot{\Phi} = \left.\frac{\partial \Phi}{\partial t}\right|_{x_1', x_2} - \dot{a}\left.\frac{\partial \Phi}{\partial x_1}\right|_{x_2, t} \tag{11}$$

and

$$\left.\frac{\partial \Phi}{\partial a}\right|_{x_1, x_2} = \left.\frac{\partial \Phi}{\partial a}\right|_{x_1', x_2} - \left.\frac{\partial \Phi}{\partial x_1}\right|_{x_2, t} \tag{12}$$

Therefore, $\dot{u}_i$ in the first part of the right-hand side of Eq. (1) can be expressed as

$$\dot{u}_i = \left.\frac{\partial u_i}{\partial t}\right|_{x_1', x_2} - \dot{a}\left.\frac{\partial u_i}{\partial x_1}\right|_{x_2, t} \tag{13}$$

Using Eq. (5) and the Reynolds transport theorem, the second part of the right-hand side in Eq. (1) becomes

$$\int_A \sigma_{ij} \dot{\varepsilon}_{ij} \, dA = \int_{A_{sta}} \frac{d\hat{w}}{dt} dA = \frac{d}{dt}\int_{A_{sta}} \hat{w} dA = \frac{d}{dt}\int_{A_{mov}} \hat{w} dA - \dot{a}\int_{\Gamma} \hat{w} n_1 d\Gamma \tag{14}$$

The proof of the Reynolds transport theorem can be found in Appendix A.

Combining Eqs. (1) (2) (13) and (14), the power balance for crack propagation can be rewritten as

$$G_s \dot{a} = \int_{\Gamma} n_j \sigma_{ij} \left(\left.\frac{\partial u_i}{\partial t}\right|_{x_1', x_2} - \dot{a}\left.\frac{\partial u_i}{\partial x_1}\right|_{t, x_2}\right) d\Gamma - \frac{d}{dt}\int_{A_{mov}} \hat{w} dA + \dot{a}\int_{\Gamma} \hat{w} n_1 d\Gamma$$

$$= \dot{a}\int_{\Gamma}\left(\hat{w} n_1 - n_j \sigma_{ij} \left.\frac{\partial u_i}{\partial x_1}\right|_{t, x_2}\right) d\Gamma + \left(\int_{\Gamma} n_j \sigma_{ij} \left.\frac{\partial u_i}{\partial t}\right|_{x_1', x_2} d\Gamma - \frac{d}{dt}\int_{A_{mov}} \hat{w} dA\right) \tag{15}$$

Therefore, the surface-forming ERR is

$$G_s = \int_{\Gamma}\left(\hat{w} n_1 - n_j \sigma_{ij} \left.\frac{\partial u_i}{\partial x_1}\right|_{t, x_2}\right) d\Gamma + \frac{1}{\dot{a}}\left(\int_{\Gamma} n_j \sigma_{ij} \left.\frac{\partial u_i}{\partial t}\right|_{x_1', x_2} d\Gamma - \frac{d}{dt}\int_{A_{mov}} \hat{w} dA\right) \tag{16}$$

Rewriting Eq. (16), we can obtain the first formula of the surface-forming EER $G_s$

**Formula I**: 
$$G_s = \int_\Gamma \left( \hat{w} n_1 - n_j \sigma_{ij} \frac{\partial u_i}{\partial x_1} \bigg|_{t,x_2} \right) d\Gamma + \left( \int_\Gamma n_j \sigma_{ij} \frac{\partial u_i}{\partial a} \bigg|_{x_1',x_2} d\Gamma - \frac{d}{da} \int_{A_{mov}} \hat{w} dA \right) \quad (17)$$

We define the first part in Eq. (17) as the crack surface $J_s$ integral, i.e.

$$J_s = \int_\Gamma (\hat{w} n_1 - n_j \sigma_{ij} \frac{\partial u_i}{\partial x_1}) d\Gamma \quad (18)$$

and the second part is denoted by

$$D \stackrel{\Delta}{=} \left( \int_\Gamma n_j \sigma_{ij} \frac{\partial u_i}{\partial a} \bigg|_{x_1',x_2} d\Gamma - \frac{d}{da} \int_{A_{mov}} \hat{w} dA \right) \quad (19)$$

We call $D$ the "derivative term", for it contains the derivatives with respect to the crack length.

It should be pointed out that the traditional $J$ integral is only a special case of the proposed general surface-forming ERR $G_s$. Because if $\hat{w}$ is taken as $\hat{w}_{-\infty}$ defined in Eq. (6), $\hat{w}$ will become the strain energy density for hyperelastic materials, and the derivative term $D$ will be zero. The surface-forming ERR $G_s$, the $J_s$ integral and the traditional $J$ integral are then identical. The related proofs can be found in Appendix B.

We note with interest that the second term $D$ of the general surface-forming ERR $G_s$ in Eq. (17) contains the derivative with respect to the crack length, which apparently implies that $G_s$ depends on the deformation and stress status of the next moment after a tiny crack propagation $\Delta a$. Therefore, it seems that $G_s$ in general condition is loading mode dependent, i.e. displacement-controlled and force-controlled loading lead to different $G_s$. In the following, we will prove that this is only an illusion and $G_s$ is loading mode independent.

Picking a finite contour $\Gamma$ surrounding the crack tip and an infinitesimal contour $\Gamma_{tip}$ as in Fig. 4, they are both traversed in the counterclockwise sense. The reverse of $\Gamma_{tip}$ in the clockwise sense is denoted by $\Gamma_{tip}^-$. Connecting $\Gamma$ and $\Gamma_{tip}^-$ by two auxiliary contours which are tight around the crack surfaces, as denoted by $\Gamma_+$ and $\Gamma_-$. $A$ is the enclosed area surrounded by $\Gamma$, and $A_{tip}$ is surrounded by the infinitesimal contour $\Gamma_{tip}$.

The expression of derivative term $D$ in Eq. (19) can then be expressed as:

$$\begin{aligned} D &= \int_\Gamma n_j \sigma_{ij} \frac{\partial u_i}{\partial a} \bigg|_{x_1',x_2} d\Gamma - \frac{d}{da} \int_{A_{mov}} \hat{w} dA \\ &= \int_{\Gamma+\Gamma_++\Gamma_-+\Gamma_{tip}^-} n_j \sigma_{ij} \frac{\partial u_i}{\partial a} \bigg|_{x_1',x_2} d\Gamma - \int_{\Gamma_++\Gamma_-} n_j \sigma_{ij} \frac{\partial u_i}{\partial a} \bigg|_{x_1',x_2} d\Gamma \\ &+ \int_{\Gamma_{tip}} n_j \sigma_{ij} \frac{\partial u_i}{\partial a} \bigg|_{x_1',x_2} d\Gamma - \frac{d}{da} \int_{A_{mov}-A_{tip}^{mov}} \hat{w} dA - \frac{d}{da} \int_{A_{tip}^{mov}} \hat{w} dA \end{aligned} \quad (20)$$

It is assumed that no external force is applied on the crack surfaces within the contour, then $n_j \sigma_{ij} = 0$ on the contour $\Gamma_+$ and $\Gamma_-$, so

$$\int_{\Gamma_+ + \Gamma_-} n_j \sigma_{ij} \frac{\partial u_i}{\partial a}\bigg|_{x_1', x_2} d\Gamma = 0 \tag{21}$$

For an observer moving with the crack tip, there should be no very dramatic change during the crack propagation of $\Delta a$. Also considering $\sigma_{ij} u_i$ is on the order of $r^0$ ($r$ is the distance to the crack tip), for an infinitesimal contour $\Gamma_{tip}$, we have

$$\int_{\Gamma_{tip}} n_j \sigma_{ij} \frac{\partial u_i}{\partial a}\bigg|_{x_1', x_2} d\Gamma = 0 \tag{22}$$

$\int_{A_{tip}^{mov}} \hat{w} dA$ should also tend to zero for an infinitesimal contour $\Gamma_{tip}$, otherwise there will be infinite average accumulated work density which is physically unreasonable. For example, the singularity of the strain energy density (i.e. the special case of $\hat{w}$) for HRR field and K field is on the order of $r^{-1}$ such that $\lim_{A_{tip}^{mov} \to 0} \int_{A_{tip}^{mov}} \hat{w} dA \to 0$. Noting that the area $A_{tip}^{mov}$ and the crack extension $\Delta a$ are two independent infinitesimal quantities, we then obtain

$$\frac{d}{da} \int_{A_{tip}^{mov}} \hat{w} dA = 0 \tag{23}$$

As a result, the derivative term Eq. (20) becomes

$$D = \int_{\Gamma + \Gamma_+ + \Gamma_- + \Gamma_{tip}^-} n_j \sigma_{ij} \frac{\partial u_i}{\partial a}\bigg|_{x_1', x_2} d\Gamma - \frac{d}{da} \int_{A^{mov} - A_{tip}^{mov}} \hat{w} dA \tag{24}$$

Utilizing the Gauss formula,

$$D = \int_{A - A_{tip}} \sigma_{ij} \frac{\partial \varepsilon_{ij}}{\partial a}\bigg|_{x_1', x_2} dA - \int_{A - A_{tip}} \frac{\partial \hat{w}}{\partial a}\bigg|_{x_1', x_2} dA$$

$$= \int_{A - A_{tip}} \left[ \sigma_{ij} \frac{\partial \varepsilon_{ij}}{\partial a}\bigg|_{x_1', x_2} - \frac{\partial \hat{w}}{\partial a}\bigg|_{x_1', x_2} \right] dA \tag{25}$$

According to Eq. (12),

$$\sigma_{ij} \frac{\partial \varepsilon_{ij}}{\partial a}\bigg|_{x_1', x_2} = \sigma_{ij} \frac{\partial \varepsilon_{ij}}{\partial a}\bigg|_{x_1, x_2} + \sigma_{ij} \frac{\partial \varepsilon_{ij}}{\partial x_1}\bigg|_{x_2, t} \tag{26}$$

$$\frac{\partial}{\partial a} \hat{w}(x_1', x_2, a) = \frac{\partial \hat{w}}{\partial a}\bigg|_{x_1, x_2} + \frac{\partial \hat{w}}{\partial x_1}\bigg|_{x_2, t} \tag{27}$$

Substituting these two equations into Eq. (25) yields

$$D = \int_{A-A_{tip}} \left( \sigma_{ij} \frac{\partial \varepsilon_{ij}}{\partial a}\bigg|_{x_1,x_2} - \frac{\partial \hat{w}}{\partial a}\bigg|_{x_1,x_2} \right) + \left( \sigma_{ij} \frac{\partial \varepsilon_{ij}}{\partial x_1}\bigg|_{x_2,t} - \frac{\partial \hat{w}}{\partial x_1}\bigg|_{x_2,t} \right) dA \qquad (28)$$

Using Eq. (5) we can simplify

$$\sigma_{ij} \frac{\partial \varepsilon_{ij}}{\partial a}\bigg|_{x_1,x_2} - \frac{\partial \hat{w}}{\partial a}\bigg|_{x_1,x_2}$$
$$= \frac{dt}{da} \left( \sigma_{ij} \frac{\partial \varepsilon_{ij}}{\partial t}\bigg|_{x_1,x_2} - \frac{\partial \hat{w}}{\partial t}\bigg|_{x_1,x_2} \right) = 0 \qquad (29)$$

Substituting Eq. (28) and Eq. (29) into Eq. (17), the second formula of the surface-forming ERR can be obtained,

**Formula II**: $\quad G_s = \int_{\Gamma} \left( \hat{w} n_1 - n_j \sigma_{ij} \frac{\partial u_i}{\partial x_1} \right) d\Gamma + \int_{A-A_{tip}} \left( \sigma_{ij} \frac{\partial \varepsilon_{ij}}{\partial x_1} - \frac{\partial \hat{w}}{\partial x_1} \right) dA \qquad (30)$

We can clearly see that $G_s$ is expressed only by the current state and deformation history, and does not contain the derivative with respect to the crack length. Therefore, the surface-forming ERR is independent on the loading mode.

Brust and Atluri (1986) have proposed a similar path-independent integral

$$T^* = \int_{\Gamma} \left( W n_1 - n_j \sigma_{ij} \frac{\partial u_i}{\partial x_1} \right) d\Gamma - \int_{A-A_{tip}} \left( \sigma_{ij} \frac{\partial \varepsilon_{ij}}{\partial x_1} - \frac{\partial W}{\partial x_1} \right) dA \qquad (31)$$

where $W$ is the total accumulated increments of stress working density and satisfies $\sigma_{ij} = \frac{\partial W}{\partial \varepsilon_{ij}}$ as in the traditional $J$ integral. We notice that the sign of the second integral is in contrast to our formula, which is a mistake in our opinion and will be explained as follows. We rewrite Eq. (30) and Eq. (31)

$$G_s = \left( \int_{\Gamma} \hat{w} n_1 d\Gamma - \int_{A-A_{tip}} \frac{\partial \hat{w}}{\partial x_1} dA \right) - \left( \int_{\Gamma} n_j \sigma_{ij} \frac{\partial u_i}{\partial x_1} d\Gamma - \int_{A-A_{tip}} \sigma_{ij} \frac{\partial \varepsilon_{ij}}{\partial x_1} dA \right) \qquad (32)$$

$$T^* = \left( \int_{\Gamma} W n_1 d\Gamma + \int_{A-A_{tip}} \frac{\partial W}{\partial x_1} dA \right) - \left( \int_{\Gamma} n_j \sigma_{ij} \frac{\partial u_i}{\partial x_1} d\Gamma + \int_{A-A_{tip}} \sigma_{ij} \frac{\partial \varepsilon_{ij}}{\partial x_1} dA \right) \qquad (33)$$

If there is no singular point in the enclosed contour, both bracket terms in Eq. (32) become zero according to the Gauss formula, and this is reasonable, while those in $T^*$ do not vanish.

The accumulated work density $\hat{w}$ in Formula II of $G_s$ (Eq. (30)) depends on the deformation history, which is not convenient for use. Considering that the former derivation process is applicable to arbitrary selection of $\hat{w}$, we can intentionally choose the current moment $t_0$ as $t_{ref}$, and set $\hat{w}_{ref}(x_1, x_2) = 0$. Then the accumulated work density $\hat{w}(x_1, x_2, t_0) = 0$ for the current moment. Applying it to Eq. (30), the third formula of the surface-forming ERR $G_s$ can be obtained

**Formula III**:
$$G_s = \int_\Gamma -n_j \sigma_{ij} \frac{\partial u_i}{\partial x_1} d\Gamma + \int_{A-A_{tip}} \sigma_{ij} \frac{\partial \varepsilon_{ij}}{\partial x_1} dA \tag{34}$$

Formula III is more concise, and is only expressed in terms of the deformation and stress status at the current moment. Therefore, we recommend using this formula to compute the surface-forming ERR $G_s$, and it is applicable for any material constitutive behaviors, such as loading/unloading in elastic-plastic crack propagation.

By comparing Eq. (30) and Eq. (34) and noting that the surface-forming ERR $G_s$ is an objective quantity, we can know that the following relationship should be met for arbitrary $\hat{w}$

$$\int_\Gamma \hat{w} n_1 d\Gamma + \int_{A-A_{tip}} -\frac{\partial \hat{w}}{\partial x_1} dA = 0 \tag{35}$$

Further simplifying this equation yields

$$\begin{aligned}
&\int_\Gamma \hat{w} n_1 d\Gamma + \int_{A-A_{tip}} -\frac{\partial \hat{w}}{\partial x_1} dA \\
&= \int_{\Gamma+\Gamma_++\Gamma_-+\Gamma_{tip}} \hat{w} n_1 d\Gamma + \int_{A-A_{tip}} -\frac{\partial \hat{w}}{\partial x_1} dA + \int_{\Gamma_{tip}} \hat{w} n_1 d\Gamma \\
&= \int_{A-A_{tip}} \frac{\partial \hat{w}}{\partial x_1} dA + \int_{A-A_{tip}} -\frac{\partial \hat{w}}{\partial x_1} dA + \int_{\Gamma_{tip}} \hat{w} n_1 d\Gamma \\
&= \int_{\Gamma_{tip}} \hat{w} n_1 d\Gamma = 0
\end{aligned} \tag{36}$$

It should be pointed out that the derivation of this interesting equation does not include any assumption and is rigorously valid. To demonstrate its correctness, the stress distribution near a crack tip of linear elastic material is used as a testing example (refer to Appendix C).

*Infinitesimal contour case*

If an infinitesimal contour $\Gamma_{tip}$ is adopted, we can further simplify the expression of $G_s$. Using Eq. (36) and noting that $A$ coincides with $A_{tip}$, Eq. (30) becomes

**Formula IV**:
$$G_s = \int_{\Gamma_{tip}} -n_j \sigma_{ij} \frac{\partial u_i}{\partial x_1} d\Gamma \tag{37}$$

which is the fourth formula of the surface-forming ERR $G_s$. We also recommend using this formula if an infinitesimal contour is adopted.

## 2.3 Formula of the surface-forming energy release rate for steady crack propagation

Steady crack propagation is a special case of elastic-plastic crack propagation. We can use its distinctive feature to simplify the general analysis. In Formula I, we adopt the accumulated work density $\hat{w}_{-\infty}$ to ensure that all material points have a unified reference state independent on the crack propagation. Steady crack propagation means that an observer moving with the crack tip cannot detect any change of physical quantities, so both $\left.\frac{\partial u_i}{\partial a}\right|_{x_1',x_2}$ and $\frac{d}{da}\int_{A_{mov}} \hat{w}_{-\infty} dA$ in Formula I vanish. Then we can obtain the fifth formula of the surface-forming ERR for steady-state crack propagation

**Formula V**: 
$$G_s = J_s = \int_{\Gamma_{tip}} \left( \hat{w}_{-\infty} n_1 - n_j \sigma_{ij} \frac{\partial u_i}{\partial x_1} \right) d\Gamma \tag{38}$$

It is easy to know that Formula V is also applicable to hyperelastic case and plastic case when unloading is absent, since $\hat{w}_{-\infty}$ degenerates into the strain energy density $w$. We recommend this formula for these situations as well.

All the formulae of the surface-forming ERR $G_s$ can be summarized in Table 1. Once $G_s$ is calculated by one of these formulae, $G_s \gtrless G_{sc}$ can be used as the criterion for crack growth.

**Table 1** Summary on all formulae of the surface-forming ERR $G_s$.

| Formula | Information needed | Applicability |
|---|---|---|
| Formula I: $G_s = \int_\Gamma \left( \hat{w} n_1 - n_j \sigma_{ij} \frac{\partial u_i}{\partial x_1} \right) d\Gamma$ $+ \left( \int_\Gamma n_j \sigma_{ij} \left.\frac{\partial u_i}{\partial a}\right|_{x_1',x_2} d\Gamma - \frac{d}{da}\int_{A_{mov}} \hat{w} dA \right)$ | ●Stress and deformation of current moment <br> ●Deformation history <br> ●Subsequent loading mode | no limitation |
| Formula II: $G_s = \int_\Gamma \left( \hat{w} n_1 - n_j \sigma_{ij} \frac{\partial u_i}{\partial x_1} \right) d\Gamma$ $+ \int_{A-A_{tip}} \left( \sigma_{ij} \frac{\partial \varepsilon_{ij}}{\partial x_1} - \frac{\partial \hat{w}}{\partial x_1} \right) dA$ | ●Stress and deformation of current moment <br> ●Deformation history | no limitation |
| Formula III (**recommended**) $G_s = \int_\Gamma \left( -n_j \sigma_{ij} \frac{\partial u_i}{\partial x_1} \right) d\Gamma$ $+ \int_{A-A_{tip}} \sigma_{ij} \frac{\partial \varepsilon_{ij}}{\partial x_1} dA$ | ●Stress and deformation of current moment | no limitation |

| Formula VI (**recommended**) $G_s = \int_{\Gamma_{tip}} \left( -n_j \sigma_{ij} \frac{\partial u_i}{\partial x_1} \right) d\Gamma$ | ● Stress and deformation of current moment | infinitesimal contour, no limitation on material behaviors |
|---|---|---|
| Formula V (**recommended**) $G_s = J_s = \int_{\Gamma} \left( \hat{w}_{-\infty} n_1 - n_j \sigma_{ij} \frac{\partial u_i}{\partial x_1} \right) d\Gamma$ | ● Stress and deformation of current moment | Hyperelastic materials, materials without plastic unloading |
| | ● Stress and deformation of current moment <br> ● Deformation history | steady-state crack propagation |

## 3. The path-independence of $G_s$ and its constituent terms

Through the definition of the surface-forming ERR, Eq. (1) and Eq. (2), it is easy to know that $G_s$ is an objective quantity with clear physical meaning, and should not depend on the contour selection by different researchers. Therefore, the path-independence of $G_s$ can be understood. In this section, we will directly prove this feature, and investigate the path-independence of $J_s$ integral and the derivative term $D$.

### 3.1 Proof for the path-independence of $G_s$ under general situations

For general elastic-plastic crack propagation problems, we investigate Formula II as an example to illustrate the path-independence of $G_s$. Consider the enclosed contour $\Gamma = \Gamma_{ii} + \Gamma_+ + \Gamma_i^- + \Gamma_-$ surrounding the crack tip in Fig. 5. $\Gamma_i^-$ is the reverse curve of $\Gamma_i$. $\Gamma_+$ and $\Gamma_-$ are tight around the crack surfaces. The area surrounded by $\Gamma_i$ and $\Gamma_{ii}$ are $A_i$ and $A_{ii}$ respectively. The surface-forming ERR of $\Gamma_i$ and $\Gamma_{ii}$ can be given in Formula II as

$$G_s^i = \int_{\Gamma_i} \left( \hat{w} n_1 - n_j \sigma_{ij} \frac{\partial u_i}{\partial x_1} \right) d\Gamma + \int_{A_i - A_{tip}} \left( \sigma_{ij} \frac{\partial \varepsilon_{ij}}{\partial x_1} - \frac{\partial \hat{w}}{\partial x_1} \right) dA \tag{39}$$

$$G_s^{ii} = \int_{\Gamma_{ii}} \left( \hat{w} n_1 - n_j \sigma_{ij} \frac{\partial u_i}{\partial x_1} \right) d\Gamma + \int_{A_{ii} - A_{tip}} \left( \sigma_{ij} \frac{\partial \varepsilon_{ij}}{\partial x_1} - \frac{\partial \hat{w}}{\partial x_1} \right) dA \tag{40}$$

The crack surfaces are assumed traction free, so $n_j \sigma_{ij} = 0$ and $n_1 = 0$ on the curve $\Gamma_+$, $\Gamma_-$. The difference between $G_s^{ii}$ and $G_s^i$ is

$$\begin{aligned} G_s^{ii} - G_s^i &= \int_{\Gamma_{ii} + \Gamma_+ + \Gamma_i^- + \Gamma_-} \left( \hat{w} n_1 - n_j \sigma_{ij} \frac{\partial u_i}{\partial x_1} \right) d\Gamma + \int_{A_{ii} - A_i} \left( \sigma_{ij} \frac{\partial \varepsilon_{ij}}{\partial x_1} - \frac{\partial \hat{w}}{\partial x_1} \right) dA \\ &= \int_{\Gamma} \left( \hat{w} n_1 - n_j \sigma_{ij} \frac{\partial u_i}{\partial x_1} \right) d\Gamma + \int_{A_{ii} - A_i} \left( \sigma_{ij} \frac{\partial \varepsilon_{ij}}{\partial x_1} - \frac{\partial \hat{w}}{\partial x_1} \right) dA \end{aligned} \tag{41}$$

There is no singular point in the area $A_{ii} - A_i$, then we can use the Gauss formula to

obtain

$$\int_\Gamma \hat{w} n_1 d\Gamma = \int_{A_{ii} - A_i} \hat{w}_{,1} dA \qquad (42)$$

and

$$\begin{aligned}
\int_\Gamma n_j \sigma_{ij} \frac{\partial u_i}{\partial x_1} d\Gamma &= \int_{A_{ii} - A_i} \left( \sigma_{ij} \frac{\partial u_i}{\partial x_1} \right)_{,j} dA \\
&= \int_{A_{ii} - A_i} \left( \sigma_{ij,j} \frac{\partial u_i}{\partial x_1} + \sigma_{ij} \frac{\partial u_{i,j}}{\partial x_1} \right) dA \\
&= \int_{A_{ii} - A_i} \sigma_{ij} \frac{\partial u_{i,j}}{\partial x_1} dA = \int_{A_{ii} - A_i} \sigma_{ij} \frac{\partial \varepsilon_{ij}}{\partial x_1} dA
\end{aligned} \qquad (43)$$

In the above derivation, no body force is assumed, i.e., $\sigma_{ij,j} = 0$.

Substituting Eqs. 错误!未找到引用源。 and (43) into Eq. (41) yields

$$G_s^{ii} - G_s^i = \int_{A_{ii} - A_i} \left( \hat{w}_{,1} - \sigma_{ij} \frac{\partial \varepsilon_{ij}}{\partial x_1} \right) dA + \int_{A_{ii} - A_i} \left( \sigma_{ij} \frac{\partial \varepsilon_{ij}}{\partial x_1} - \frac{\partial \hat{w}}{\partial x_1} \right) dA = 0 \qquad (44)$$

The path-independence of $G_s$ is then directly proven.

## 3.2 Investigation on the path-independence of $J_s$ integral and $D$

In Formula I and Formula II, $G_s = J_s + D$. Here we further investigate whether $J_s$ integral and $D$ are path-independent or not. It is found that under general situations, their path-independence cannot be proven analytically. As we have known that $G_s$ is path-independent, so $J_s$ and $D$ can only be path dependent or path-independent at the same time. Here we use the following example to demonstrate that $J_s$ is path dependent under elastic-plastic unsteady-state crack growth.

As shown in Fig. 6, there exists a crack in the middle of a strip, whose upper and lower boundaries are subjected to constant displacement loadings. For the two regions $R^i$ and $R^{ii}$ far ahead of the crack tip, we may artificially make them have different plastic deformation histories by pre-loading and unloading while keeping their final stress and deformation state the same as neighboring regions, such that the $\hat{w}$ in these two regions are different. We choose two contours, $\Gamma_i = \Gamma_1 + \Gamma_2 + \Gamma_3^i + \Gamma_4 + \Gamma_5$ and $\Gamma_{ii} = \Gamma_1 + \Gamma_2 + \Gamma_3^{ii} + \Gamma_4 + \Gamma_5$. $\Gamma_3^i$ goes through the region $R^i$ and $\Gamma_3^{ii}$ goes through the region $R^{ii}$ as in Fig. 6.

We then calculate the difference between $J_s^{\text{i}}$ and $J_s^{\text{ii}}$. As the upper and lower boundaries $\Gamma_4$, $\Gamma_2$ are fixed, $n_1 = 0$, $\frac{\partial u_i}{\partial x_1} = 0$, so they have no contribution to the value of $J_s$. The integral $J_s^{\text{i}}$ and $J_s^{\text{ii}}$ have the same value on $\Gamma_1$, $\Gamma_5$, so we only focus on the curves $\Gamma_3^{\text{i}}$ and $\Gamma_3^{\text{ii}}$. Assuming they are far away from the crack tip, then $n_1 = 1$ and $\frac{\partial u_i}{\partial x_1} = 0$ hold for these two curves. We can have

$$J_s^{\text{i}} = J_s^{\Gamma_1} + J_s^{\Gamma_5} + \int_{\Gamma_3^{\text{i}}} \hat{w} d\Gamma \tag{45}$$

$$J_s^{\text{ii}} = J_s^{\Gamma_1} + J_s^{\Gamma_5} + \int_{\Gamma_3^{\text{ii}}} \hat{w} d\Gamma \tag{46}$$

Considering that the accumulated work density $\hat{w}$ are different on $\Gamma_3^{\text{i}}$ and $\Gamma_3^{\text{ii}}$ due to the artificially preloading, as mentioned above, we obtain

$$J_s^{\text{i}} \neq J_s^{\text{ii}} \tag{47}$$

Therefore, $J_s$ integral and $D$ are usually path dependent. However, for a crack in hyperelastic materials, or a steady-state propagating crack and a non-propagating crack in elastic-plastic materials, $J_s$ possesses the path-independence since $G_s = J_s$ as discussed above.

## 4. Physical interpretations and discussions

### 4.1 The accumulated work density vs. the strain energy density

In traditional J integral, the strain energy density

$$w(\varepsilon_{ij}) = \int_0^{\varepsilon_{ij}} \sigma_{ij} d\varepsilon_{ij} \tag{48}$$

is used for hyperelastic materials or elastic-plastic materials without plastic unloading, and is a single valued function of the strain. It has a clear physical meaning, representing the underneath area of the stress-strain curve ( as shown in Fig. 7a) or the stored energy. However, once plastic unloading appears, this definition of the strain energy density is not a single valued function of the strain any more, and also depends on the loading paths. Therefore, the J integral loses its applicability as presented in Appendix B. In addition, since the strain $\varepsilon_{ij}$ is an integration variable in Eq. (48), it is supposed to increase monotonically, which makes this definition unable to reflect irreversible loading and unloading process.

The accumulated work density $\hat{w}_{-\infty}(t) = \int_{-\infty}^{t} \sigma_{ij} \dot{\varepsilon}_{ij} d\tau$, defined as the accumulation of the power by the internal force, however, always has a clear and physically reasonable definition for complicated loading/unloading process, since the monotonically increasing time is used as the integration variable. The physical meaning of $\hat{w}_{-\infty}$ can also be shown by the area under the stress-strain curve ( see Fig. 7b and c).

Obviously, the strain energy density $w$ is a special case of the accumulated work density $\hat{w}_{-\infty}$, therefore the latter has wider applicability.

### 4.2 $G_s$ vs. $G$ for steady-state crack propagation in elastic-plastic materials

In this subsection, we will explain with examples why the surface-forming ERR $G_s$ is superior to the traditional ERR $G$. Figure 8a shows an infinite strip of height $h$ with a straight crack located in the middle. The upper and lower surfaces of the strip are clamped so that the displacement is constant on these two boundaries. $\Gamma = \Gamma_1 + \Gamma_2 + \Gamma_3 + \Gamma_4 + \Gamma_5$ denoted by the dashed line is taken as the integral contour. $\Gamma_1$ and $\Gamma_5$ are far behind of the crack tip and $\Gamma_3$ is far ahead of the crack tip. For steady-state crack propagation in elastic-plastic materials, one of the widely accepted method to calculate the energy release rate $G$ is subtracting the energy density of $\Gamma_1$ and $\Gamma_5$ from that of $\Gamma_3$. Here the energy is defined as the ability to do external work. Since $\Gamma_1$ and $\Gamma_5$ are approximately stress free, it is usually considered that their energy density is zero. The entire contribution to $G$ comes only from $\Gamma_3$ and

$$G = w_{\Gamma_3} h \tag{49}$$

where $w_{\Gamma_3}$ is the constant strain energy density at $\Gamma_3$. Obviously, the traditional energy release rate $G$ includes both the energy used to separate the crack surfaces (or surface-forming energy) and the accompanying plastic dissipation away the crack surfaces. In elastic-plastic crack propagation, the plastic strain and yielding zone usually depend on the loading conditions. For example, steady-state crack propagation can be realized in a strip specimen by applying displacement loading as shown in Fig. 8a or moments at the left ends as shown in Fig. 8b, but their plastic zones and plastic dissipation energy should be different. Therefore, the corresponding critical ERR $G_c$ for steady-state crack propagation is loading mode dependent and is not a material constant.

However, by computing the difference between the accumulated work density $\hat{w}_{-\infty}$ at $\Gamma_3$ and $\hat{w}_{-\infty}$ at $\Gamma_1$ and $\Gamma_5$, as shown in Fig. 8a, the surface-forming ERR $G_s$ excludes the loading-mode-dependent plastic dissipation, and the corresponding fracture criterion $G_s \geq G_{sc}$ has wider applicability.

### 4.3 The surface-forming ERR $G_s$ for the cases with crack surface tractions

In previous sections, it is assumed the crack surfaces within the contour are traction free. If there are tractions, we may extend the contour towards the crack tip along the crack surfaces, as shown in Fig. 9. Then the surface-forming ERR $G_s$ (Formula III) for arbitrary contour becomes

$$\begin{aligned} G_s &= \int_{\Gamma+\Gamma_++\Gamma_-} \left(-n_j \sigma_{ij} \frac{\partial u_i}{\partial x_1}\right) d\Gamma + \int_{A-A_{tip}} \sigma_{ij} \frac{\partial \varepsilon_{ij}}{\partial x_1} dA \\ &= \left[\int_\Gamma \left(-n_j \sigma_{ij} \frac{\partial u_i}{\partial x_1}\right) d\Gamma + \int_{A-A_{tip}} \sigma_{ij} \frac{\partial \varepsilon_{ij}}{\partial x_1} dA\right] + \int_{\Gamma_++\Gamma_-} \left(-n_j \sigma_{ij} \frac{\partial u_i}{\partial x_1}\right) d\Gamma \end{aligned} \tag{50}$$

## 5. Conclusions

In this paper, through the analysis on the power balance during the crack propagation, we propose a surface-forming energy release rate $G_s$, which represents the energy directly dissipated on the surface-forming during the crack propagation and excludes the loading-mode-dependent plastic dissipation. The surface-forming ERR based fracture criterion has no limitation on the constitutive behaviors of materials, so its applicability is wider. Several formulae in integration form have been derived to calculate $G_s$ under different conditions, and the following conclusions are drawn:

- For the general elastic-plastic crack propagation cases, we recommend Formula III, which only depends on the stress and deformation of the current moment and the definition of the energy is avoided. When an infinitesimal contour is chosen, the formula becomes more concise.
- A new $J_s$ integral is also derived, and the traditional $J$ integral is only a special case of the $J_s$ integral. However, we recommend using the $G_s$ based fracture criterion, rather than the $J_s$ integral (or $J$ integral) based fracture criterion. Because the surface-forming energy release rate $G_s$ is always path independent but $J_s$ is not.
- For hyperelastic materials and steady-state crack propagation, the surface-forming ERR $G_s$ and the $J_s$ integral are identical.
- The physical meanings and applicability of the proposed surface-forming ERR, the traditional ERR, $J_s$ integral and $J$ integral are compared and discussed. It has been illustrated that the surface-forming ERR $G_s$ based criterion is more applicable than others.

Acknowledgement

The authors acknowledge the support from National Natural Science Foundation of China (Grant Nos. 11090334, 51232004 and 11372158), National Basic Research Program of China (973 Program Grant Nos. 2010CB832701) and Tsinghua University Initiative Scientific Research Program (No. 2011Z02173).

## Appendix A

As shown in Fig. A1, $A_{sta}$ represents the area surrounded by a stationary contour and $A_{mov}$ represents the area surrounded by a moving contour. At the current moment, $A_{sta}$ and $A_{mov}$ coincide. After a period time $\Delta t$, $A_{mov}$ deviates from $A_{sta}$, and there are three regions: I, II, and III. $A_{sta}$ is always composed of region I and region II, but $A_{mov}$ consists of different regions---region I and II at the current moment $t$, region II and III at the moment $t+\Delta t$.

Define the accumulated work of the area $A$ enclosed by a closed contour $\Gamma$ as:

$$\hat{U} = \int_A \hat{w} dA \tag{A1}$$

Then the accumulated work of $A_{sta}$ at the moment $t$ and $t+\Delta t$ are

$$\hat{U}_{sta}(t) = \hat{U}_I(t) + \hat{U}_{II}(t) \tag{A2}$$

$$\hat{U}_{sta}(t+\Delta t) = \hat{U}_I(t+\Delta t) + \hat{U}_{II}(t+\Delta t) \tag{A3}$$

The difference between them is

$$\Delta \hat{U}_{sta} = \hat{U}_I(t+\Delta t) - \hat{U}_I(t) + \left[\hat{U}_{II}(t+\Delta t) - \hat{U}_{II}(t)\right] \tag{A4}$$

The accumulated work of $A_{mov}$ at the moment $t$ and $t+\Delta t$ are

$$\hat{U}_{mov}(t) = \hat{U}_I(t) + \hat{U}_{II}(t) \tag{A5}$$

$$\hat{U}_{mov}(t+\Delta t) = \hat{U}_{II}(t+\Delta t) + \hat{U}_{III}(t+\Delta t) \tag{A6}$$

Then their difference is

$$\Delta \hat{U}_{mov} = \left[\hat{U}_{II}(t+\Delta t) - \hat{U}_{II}(t)\right] - \hat{U}_I(t) + \hat{U}_{III}(t+\Delta t) \tag{A7}$$

Subtracting Eq. (A7) from Eq. (A4) yields

$$\begin{aligned}\Delta \hat{U}_{sta} - \Delta \hat{U}_{mov} &= \hat{U}_I(t+\Delta t) - \hat{U}_{III}(t+\Delta t) \\ &= -\Delta t \, \dot{a} \int_{\Gamma_2} \hat{w} n_1 d\Gamma - \Delta t \, \dot{a} \int_{\Gamma_1} \hat{w} n_1 d\Gamma = -\Delta t \, \dot{a} \int_{\Gamma} \hat{w} n_1 d\Gamma\end{aligned} \tag{A8}$$

The integrations in the equation above can be easily understood from Fig. A1. Dividing both sides of Eq. (A8) by $\Delta t$, the following Reynolds transport theorem can be obtained

$$\frac{\Delta \hat{U}_{sta} - \Delta \hat{U}_{mov}}{\Delta t} = \frac{d}{dt}\int_{A_{sta}} \hat{w} dA - \frac{d}{dt}\int_{A_{mov}} \hat{w} dA = -\dot{a}\int_{\Gamma} \hat{w} n_1 d\Gamma \tag{A9}$$

**Appendix B**

For materials without plastic unloading, $\hat{w}$ becomes the strain energy density $w$ as stated before and satisfies

$$\sigma_{ij} = \frac{\partial w}{\partial \varepsilon_{ij}} \tag{B1}$$

Consider an enclosed contour $\Gamma = \Gamma_i^- + \Gamma_- + \Gamma_{ii} + \Gamma_+$ as shown in Fig. 5, the enclosed area $A = A_{ii} - A_i$. $\Gamma_i^-$ is the reverse curve of $\Gamma_i$, and $\Gamma_+, \Gamma_-$ are tight around the crack surfaces. Because there is no singular point in the area, according to the Gauss formula, the first term of the right hand side in Eq. (19) becomes

$$\int_{\Gamma} n_j \sigma_{ij} \frac{\partial u_i}{\partial t}\bigg|_{x_1', x_2} d\Gamma = \int_A \left(\sigma_{ij} \frac{\partial u_i}{\partial t}\bigg|_{x_1', x_2}\right)_{,j} dA$$
$$= \int_A \sigma_{ij} \frac{\partial u_{i,j}}{\partial t}\bigg|_{x_1', x_2} d\Gamma = \int_A \sigma_{ij} \frac{\partial \varepsilon_{ij}(x_1', x_2, t)}{\partial t} dA \tag{B2}$$

The second term becomes

$$\frac{d}{dt}\int_{A_{mov}} \hat{w} dA = \int_{A_{mov}} \frac{\partial}{\partial t} w(x_1', x_2, t) dA$$
$$= \int_{A_{mov}} \frac{\partial}{\partial t} w(\varepsilon_{ij}(x_1', x_2, t), x_2) dA = \int_{A_{mov}} \frac{\partial w}{\partial \varepsilon_{ij}} \frac{\partial \varepsilon_{ij}(x_1', x_2, t)}{\partial t} dA \tag{B3}$$
$$= \int_{A_{mov}} \sigma_{ij} \frac{\partial \varepsilon_{ij}(x_1', x_2, t)}{\partial t} dA$$

Substituting Eq. (B2) and Eq. (B3) into Eq. (19), $D = 0$ is proven. Therefore, for materials without plastic unloading, the surface-forming ERR $G_s$ equals to $J_s$ or J integrals.

However, for a propagating crack with plastic unloading, the strain energy density is not a single valued function of the strain any more, and the following relation in the derivation of Eq. (B3)

$$\int_{A_{mov}} \frac{\partial}{\partial t} w(x'_1, x_2, t) dA = \int_{A_{mov}} \frac{\partial}{\partial t} w\big(\varepsilon_{ij}(x'_1, x_2, t), x_2\big) dA$$
$$= \int_{A_{mov}} \frac{\partial w}{\partial \varepsilon_{ij}} \frac{\partial \varepsilon_{ij}(x'_1, x_2, t)}{\partial t} dA = \int_{A_{mov}} \sigma_{ij} \frac{\partial \varepsilon_{ij}(x'_1, x_2, t)}{\partial t} dA \quad (B4)$$

cannot be obtained. Then the derivative term $D$ cannot vanish, and the J integral neither represents the surface-forming ERR $G_s$ nor possesses other clear physical meaning.

**Appendix C**

For linear elastic materials, the strain energy density can be expressed as

$$\hat{w} = w = \frac{1}{2} \sigma_{ij} \varepsilon_{ij} \quad (C1)$$

The stress field in plane stress condition is (Anderson, 2005)

$$\begin{Bmatrix} \sigma_{11} \\ \sigma_{12} \\ \sigma_{22} \end{Bmatrix} = \frac{K_I}{\sqrt{2\pi r}} \cos\frac{\theta}{2} \begin{Bmatrix} 1 - \sin\frac{\theta}{2} \sin\frac{3\theta}{2} \\ \sin\frac{\theta}{2} \cos\frac{3\theta}{2} \\ 1 + \sin\frac{\theta}{2} \sin\frac{3\theta}{2} \end{Bmatrix} \quad (C2)$$

The corresponding strain field can be calculated by the following constitutive equations,

$$\begin{Bmatrix} \varepsilon_{11} \\ \varepsilon_{12} \\ \varepsilon_{22} \end{Bmatrix} = \frac{1}{E} \begin{Bmatrix} \sigma_{11} - \upsilon \sigma_{22} \\ 2(1+\upsilon)\sigma_{12} \\ \sigma_{22} - \upsilon \sigma_{11} \end{Bmatrix} \quad (C3)$$

From Eqs. (C2) and (C3), the strain energy density in Eq. (C1) becomes

$$w = \frac{K_I^2}{4\pi E r} \cos^2\frac{\theta}{2} \left[ \left(1-\sin\frac{\theta}{2}\sin\frac{3\theta}{2}\right)^2 + \left(1+\sin\frac{\theta}{2}\sin\frac{3\theta}{2}\right)^2 \\ +4(1-\upsilon)\sin^2\frac{\theta}{2}\cos^2\frac{3\theta}{2} - 2\upsilon\left(1-\sin^2\frac{\theta}{2}\sin^2\frac{3\theta}{2}\right) \right] \quad (C4)$$

Substituting the above equation into Eq. (36) yields

$$\int_{\Gamma_\varepsilon} \hat{w} n_1 d\Gamma = \lim_{r \to 0} \int_0^{2\pi} \hat{w} n_1 r d\theta$$

$$= \lim_{r \to 0} \int_0^{2\pi} \frac{K_I^2}{4\pi E} \cos\theta \cos^2\frac{\theta}{2} \left[ \begin{array}{c} 2 + 2\sin^2\frac{\theta}{2}\sin^2\frac{3\theta}{2} + 4(1-\upsilon)\sin^2\frac{\theta}{2}\cos^2\frac{3\theta}{2} \\ -2\upsilon\left(1 - \sin^2\frac{\theta}{2}\sin^2\frac{3\theta}{2}\right) \end{array} \right] d\theta \quad (C5)$$

$$= 0$$

For a plane strain case (i.e. $\varepsilon_{33} = const$), $\sigma_{33} = \upsilon(\sigma_{11} + \sigma_{22})$, and the elastic modulus $E$ changes to be $E' = \frac{E}{1-\upsilon^2}$. We can similarly obtain $\lim_{r \to 0} \int_0^{2\pi} \hat{w} n_1 r dr = 0$.

Orowan, E., 1949. Fracture and Strength of solids. Reports on Progress in Physics , 12(1),185.

Rice, J. R., 1967. A path independence integral and the approximate analysis of strain concentration by notches and cracks. Advanced Research Projects Agency. SD-86.

Simha, N. K., Fischer, F. D., Shan, G. X., Chen, C. R., Kolednik, O., 2008. J-integral and crack driving force in elastic-plastic materials. J. Mech. Phys. Solids, 56:2876~2895.

Sumpter, J. D. G., Turner, C. E., 1976. Use of the J contour integral in elastic-plastic fracture studies by finite-element methods. J. Mech. Eng. Sci. DoI: 10. 1243

Wells, A A., 1963. Application of fracture mechanics at and beyond general yielding. British Welding J. 10:563~570.

Wnuk, M. P., READ, D. T., 1986. Essential work of fracture (We) versus energy dissipation rate (Jc) in plane stress ductile fracture. Int J Fracture. 31:161-171.

Vu-Khanh, T., 1994. Impact fracture characterization of polymer with ductile behavior. Theor Appl Fract Mech. 21:83–90.21

**Figure captions**

Figure 1 Schematics of a propagating crack in elastic-plastic materials: (a) plastic yielding near a crack tip before the initiation, (b) the growth of the plastic zone accompanying the crack propagation, (c) steady-state crack propagation.

Figure 2 Schematic of a stationary contour surrounding the crack tip.

Figure 3 Stationary and moving coordinate systems and corresponding contours.

Figure 4 Schematic diagram of an enclosed integral contour including an outer contour and an infinitesimal contour closely surrounding the crack tip.

Figure 5 Schematic diagram of an enclosed contour including two different contours, $\Gamma_i$ and $\Gamma_{ii}$.

Figure 6 A cracked strip with constant displacement loading on upper and lower boundaries.

Figure 7 Schematics of the strain energy density of (a) hyperelastic materials, and the accumulated work density of elastic-plastic materials in (b) loading stage and (c) unloading stage.

Figure 8 Schematics of steady-state crack propagation in a strip under (a) constant displacement loadings and (b) constant moment loadings at the left ends.

Figure 9 Integral contour for the cases with traction on the crack surfaces.

Figure A1 Areas surrounded by a stationary contour and a moving contour.

(a)

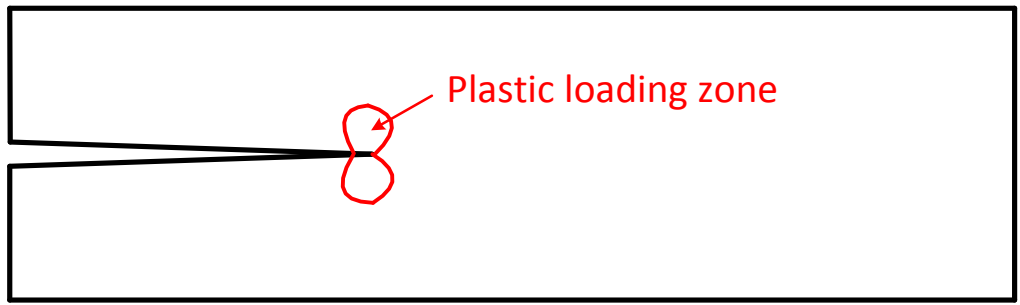

(b)

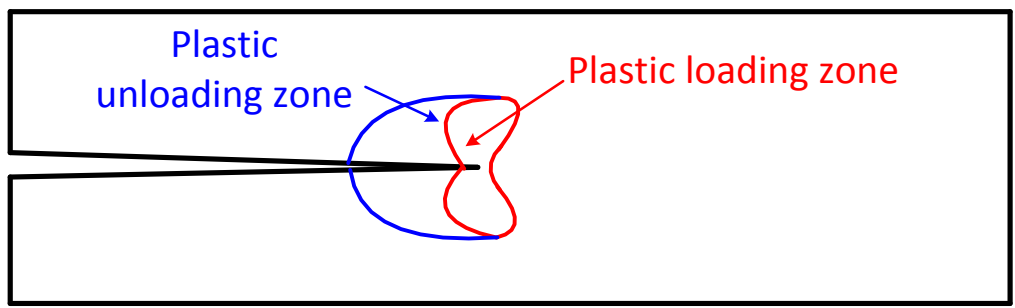

(c)

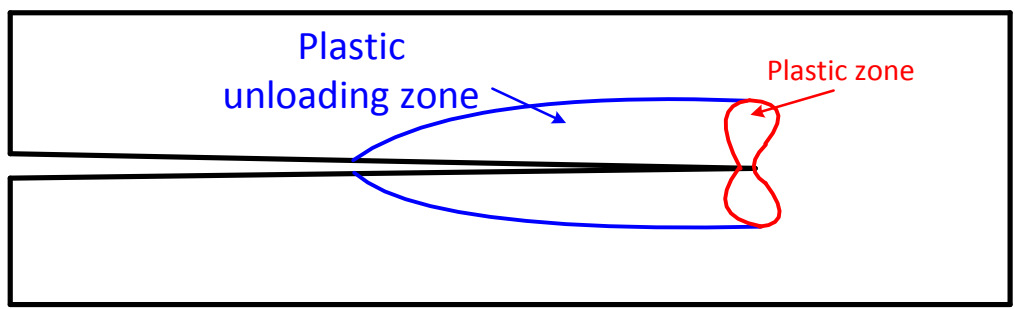

Fig. 1. Schematics of a propagating crack in elastic-plastic materials: (a) plastic yielding near a crack tip before the initiation, (b) the growth of the plastic zone accompanying the crack propagation, (c) steady-state crack propagation.

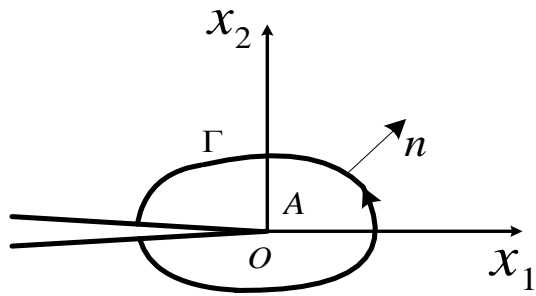

Fig. 2. Schematic of a stationary contour surrounding the crack tip

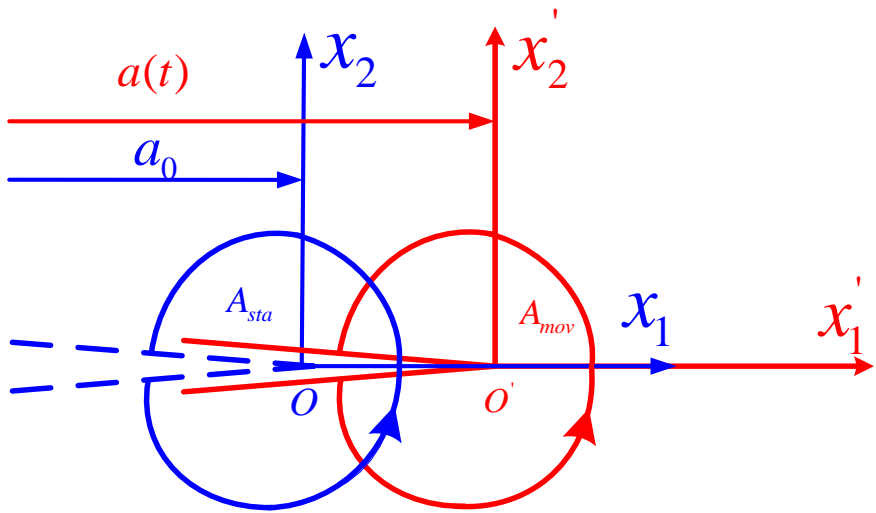

Fig. 3. Stationary and moving coordinate systems and corresponding contours

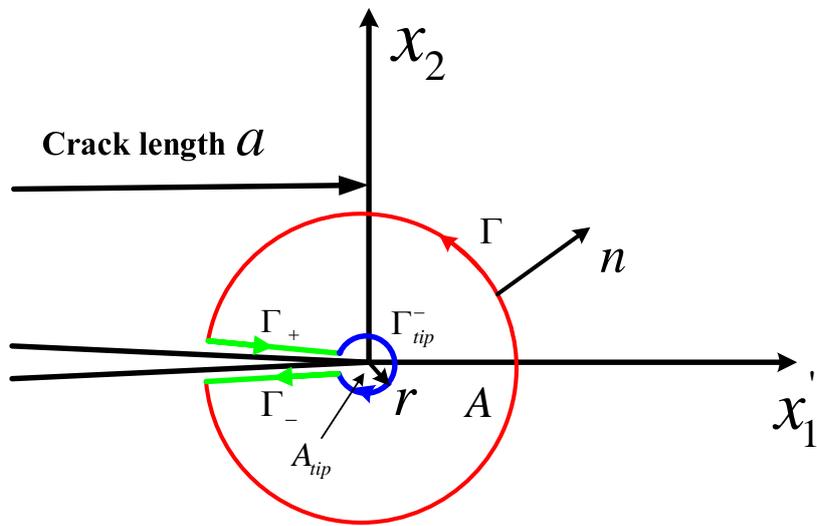

Fig. 4. Schematic diagram of an enclosed integral contour including an outer contour and an infinitesimal contour closely surrounding the crack tip

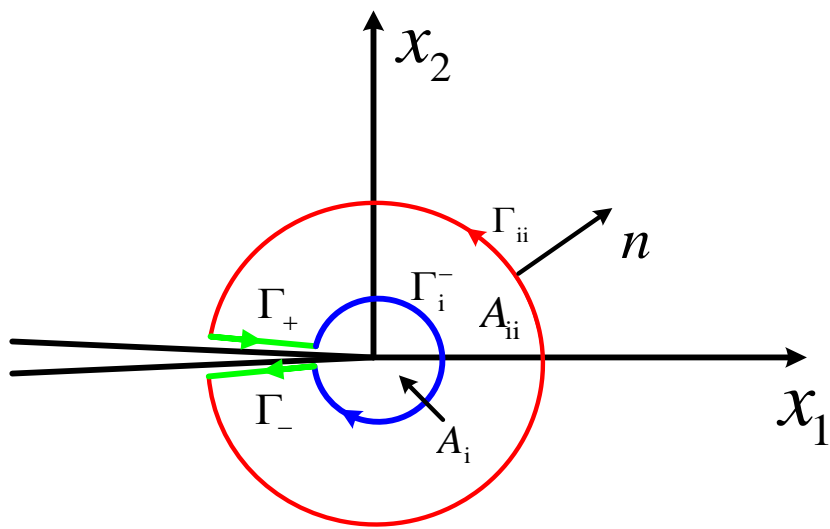

Fig. 5. Schematic diagram of an enclosed contour including two different contours, $\Gamma_i$ and $\Gamma_{ii}$

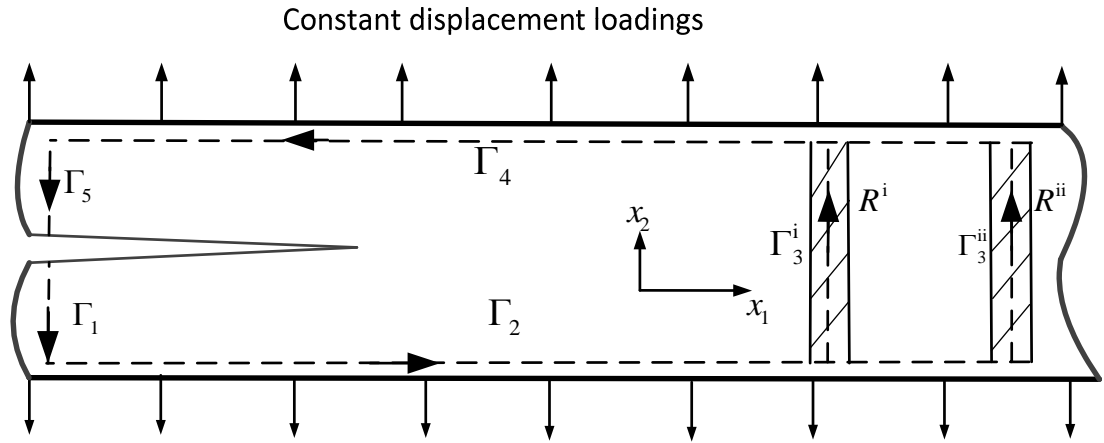

Fig. 6. A cracked strip with constant displacement loading on upper and lower boundaries

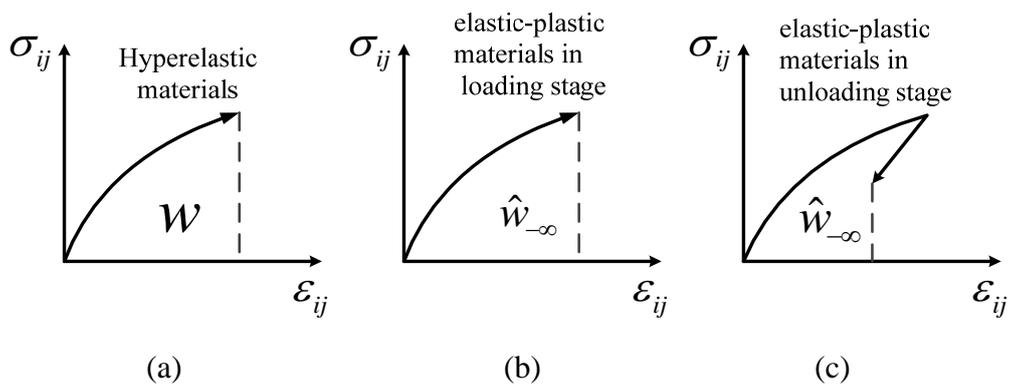

(a)                      (b)                    (c)

Fig. 7 Schematics of the strain energy density of (a) hyperelastic materials, and the accumulated work density of elastic-plastic materials in (b) loading stage and (c) unloading stage

(a):

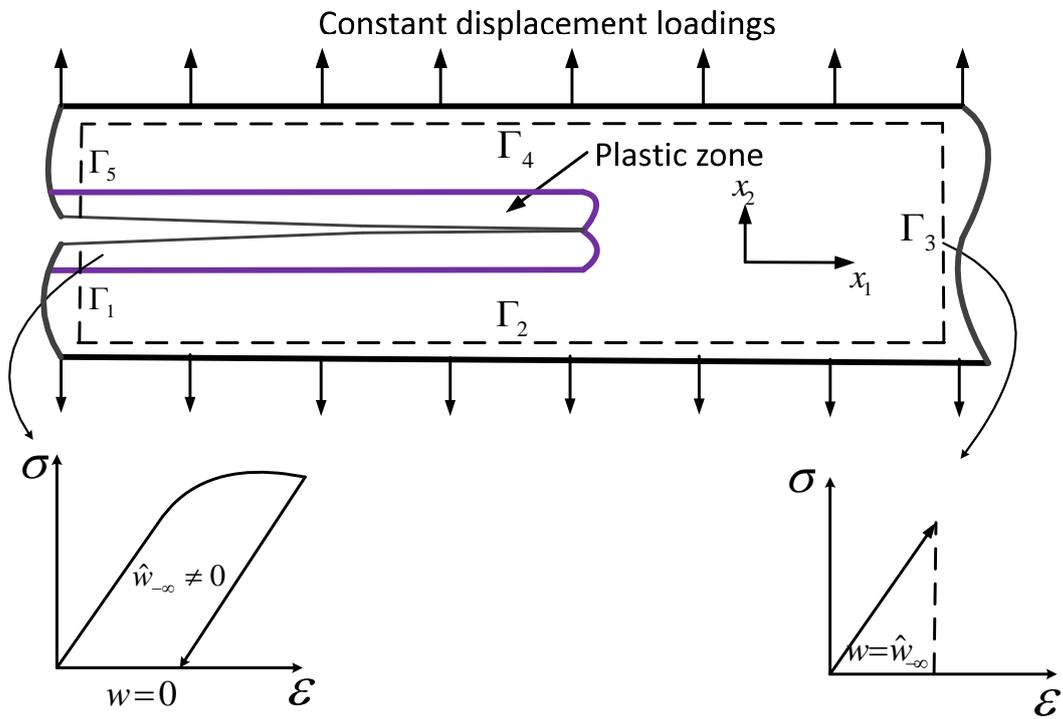

(b):

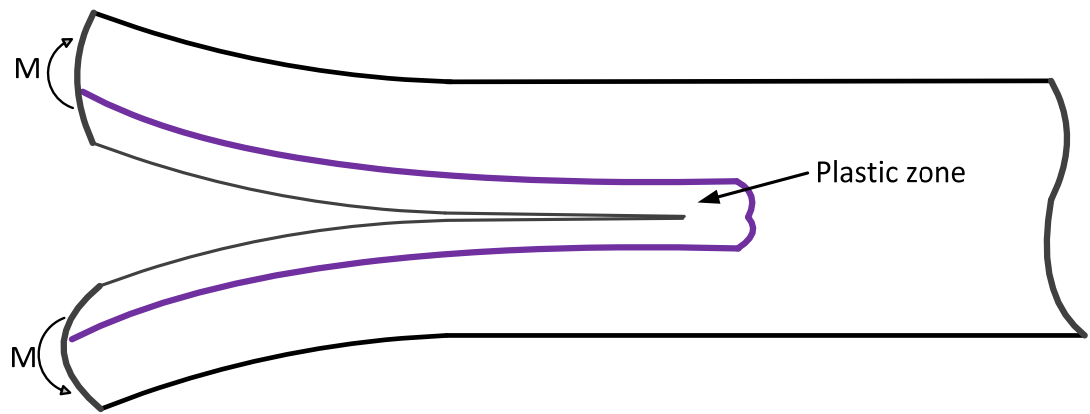

Fig. 8. Schematics of steady-state crack propagation in a strip under (a) constant displacement loadings and (b) constant moment loadings at the left ends.

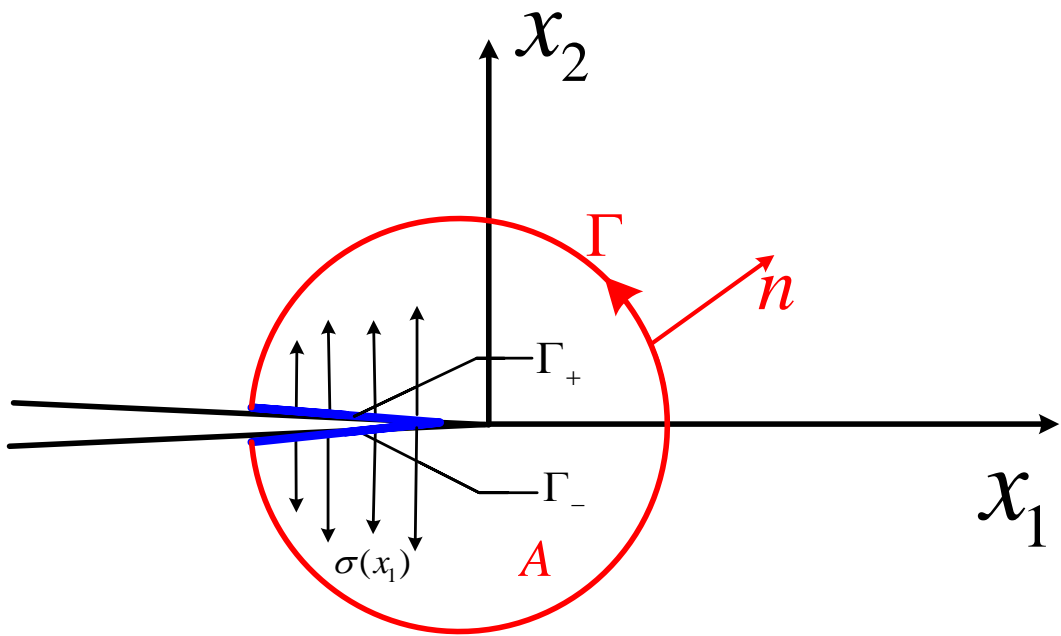

Fig. 9. Integral contour for the cases with traction on the crack surfaces

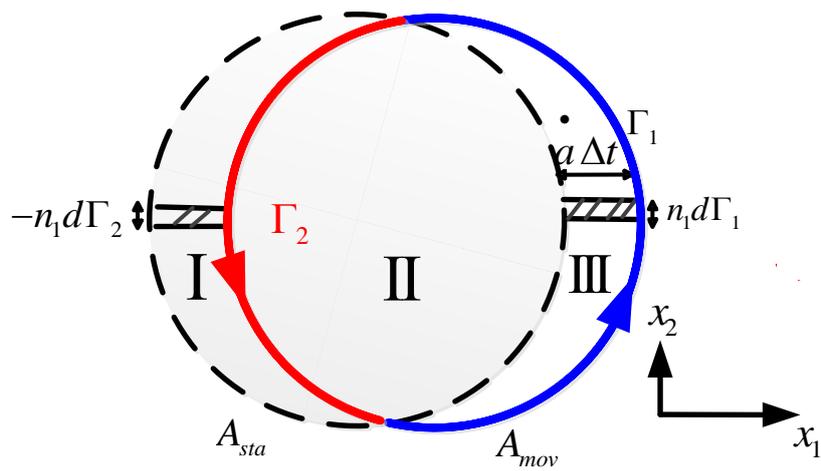

Fig. A1. Areas surrounded by a stationary contour and a moving contour